\documentclass[letterpaper, 10 pt, conference]{ieeeconf}  

\IEEEoverridecommandlockouts                              

\usepackage{amsmath, amssymb, amsfonts}
\usepackage{cite}
\usepackage{graphicx}
\usepackage{subcaption}

\newcounter{remark}
\newenvironment{remark}{%
  \refstepcounter{remark}\par\noindent\textbf{Remark~\theremark.}\ }{\par}

\title{\LARGE \bf
Cooperative Transportation Without Prior Object Knowledge via Adaptive Self-Allocation and Coordination
}

\author{Jie Song$^{1}$, Yang Bai$^{2}$ and Naoki Wakamiya$^{1}$
\thanks{$^{1}$J. Song and N. Wakamiya are with the Department of Bioinformatic Engineering, University of Osaka, Osaka, Japan
        {\tt\small j-song@ist.osaka-u.ac.jp; wakamiya@ist.osaka-u.ac.jp}}%
\thanks{$^{2}$Y. Bai is with the Graduate School of Advanced Science and Engineering, Hiroshima University, Hiroshima, Japan
        {\tt\small yangbai@hiroshima-u.ac.jp}}%
}

\begin{document}

\maketitle
\thispagestyle{empty}
\pagestyle{empty}

\begin{abstract}

This work proposes a novel cooperative transportation framework for multi-agent systems that does not require any prior knowledge of cargo locations or sizes. Each agent relies on local sensing to detect cargos, recruit nearby agents, and autonomously form a transportation team with an appropriate size. The core idea is that once an agent detects a cargo within its sensing range, it generates an attraction field represented by a density function, which pulls neighboring agents toward the cargo. When multiple cargos are present, the attraction fields generated by different agents are adaptively weighted and combined with Centroidal Voronoi Tessellation (CVT), enabling agents to self-organize into balanced formations while automatically allocating more agents to larger cargos. To prevent agents from clustering on one side of a large cargo, a Control Barrier Function (CBF)-based mechanism is introduced to enforce safe inter-agent distances and promote a uniform, symmetric distribution of agents around each cargo, which is essential for stable transportation. Simulation results demonstrate that the proposed framework can simultaneously transport multiple cargos of different sizes in a coordinated and collision-free manner.

\end{abstract}


\section{INTRODUCTION}
In recent years, cooperative robotic transportation has attracted increasing attention, driven by growing demands for transporting large-scale, heavy, and irregular objects in real-world applications \cite{Yamashita2003, Abbas2023, Vlantis2022, Izadbakhsh2022, Sieber2018}. In such scenarios, traditional single-robot grasp-and-carry approaches suffer from inherent limitations in payload capacity and system robustness, and often rely on precise object modeling, grasp point selection, and environmental perception \cite{Kemp2007, Wu2015, Iriondo2019, Thakar2020, Kim2013, ParraVega2013}. In contrast, multi-robot cooperative transportation enables load sharing among multiple agents and improves robustness against individual failures and environmental uncertainties, making it a promising paradigm for addressing complex transportation tasks \cite{Wan2017, Lin2022, Wang2023, Tejada2025, Eoh2021, Alkilabi2017, magid2022asia}.

Among existing multi-robot cooperative transportation approaches, caging-based methods represent a prominent class \cite{Wang2002ObjectClosure, Wang2003CCClosure, Zhang2019GeometricCage, Nedjah2025Caging, Su2021ImmobilizingCaging}. These methods achieve object transportation by geometrically enclosing the target object with multiple robots, rather than relying on explicit grasping. By reducing the dependence on precise grasp execution and force control, caging-based approaches offer increased robustness when handling objects with irregular shapes or uncertain contact conditions. However, despite these advantages, most existing caging-based cooperative transportation methods still rely on strong prior assumptions. In particular, the target object’s location is typically assumed to be known in advance, key physical properties such as object size and shape are either predefined or accurately estimated beforehand, and the number of robots required for transportation is often fixed a priori. Such assumptions significantly limit the applicability of existing methods in real-world scenarios, where objects may be newly discovered, partially observed, or subject to substantial uncertainty.

To address these limitations, this paper proposes a multi-agent cooperative transportation framework that operates without prior knowledge of object properties. The proposed framework does not require advance information about object location, size, or weight. Instead, agents rely solely on local sensing to autonomously detect target objects and form cooperative transportation structures through distributed interactions. During this process, the number of participating agents is adaptively determined based on local observations, without centralized coordination or pre-specified team sizes.

Specifically, a dynamically constructed density function is introduced to guide agents toward detected objects, and this density-driven mechanism is integrated with Centroidal Voronoi Tessellation (CVT) to enable self-organized spatial distribution and cooperative formation around each object. The adaptive weighting of the density function naturally results in larger objects attracting more agents for transportation. Furthermore, to prevent agents from clustering on one side of an object and to ensure safety during close-proximity cooperation, Control Barrier Functions (CBFs) are incorporated at the control level. This CBF-based mechanism explicitly enforces safe inter-agent distances and promotes a uniform and balanced distribution of agents around each object, which is critical for achieving stable cooperative transportation. Simulation results demonstrate that the proposed framework enables the simultaneous transportation of multiple objects with different sizes under unknown object conditions, validating its effectiveness in terms of adaptability, robustness, and safety.

The rest of paper is organized as follows: In Section II, some preliminaries about CVT are introduced. Section III presents the controller design, including the integration of CVT and the construction of the density function. Simulation results demonstrating the effectiveness of the proposed method are provided in Section IV, and conclusions are drawn in Section V.

\section{Preliminary} \label{Preliminary}
In this section, we briefly review CVT, which serves as the geometric foundation of our proposed coordination framework. It provides a systematic way to achieve an optimal distribution of agents over a workspace. Our control algorithm builds on this CVT structure and extends it to handle unknown objects and adaptive self-grouping.

Define $\mathcal{D} \subset \mathbb{R}^2$ as a two-dimensional domain representing the operational workspace in which $N$ agents operate. Let $\mathcal{N} = \{1, \dots, N\}$ be the index set of all agents. Let $\boldsymbol{p} = \{ \boldsymbol{p}_i \in \mathbb{R}^2 \mid i \in \mathcal{N} \}$ denote the collection of position vectors of all $N$ agents. A point $\boldsymbol{q} \in \mathcal{D}$ represents an arbitrary location within the workspace. The Voronoi tessellation over the domain is given by:
\begin{equation}\label{voronoi}
V_{i}\left ( \boldsymbol{p} \right )= \left \{ \boldsymbol{q} \in \mathcal{D} \mid \left \| \boldsymbol{q} - \boldsymbol{p}_i \right \| \leq \left \| \boldsymbol{q} - \boldsymbol{p}_j \right \|,\, \forall j \neq i \right \},
\end{equation}
ensuring that each agent $i$ is associated with a specific Voronoi subregion $V_i$.

To evaluate the effectiveness of the agent distribution, we define the locational cost function as:
\begin{equation}\label{H0}
\mathcal{H}(\boldsymbol{p}, t) = \sum_{i=1}^{N} \int_{V_i} \left \| \boldsymbol{q} - \boldsymbol{p}_i \right \|^2 \, \phi(\boldsymbol{q}, t) \, \mathrm{d} \boldsymbol{q},
\end{equation}
where $\phi(\boldsymbol{q}, t)$ is a density function, assumed to be strictly positive and bounded over the domain $\mathcal{D}$. This function represents the relative importance of different spatial locations and guides agents to concentrate in higher-density regions.

To achieve an optimal spatial distribution of agents within the domain $\mathcal{D}$ at any time $t$, the agent positions $\boldsymbol{p}$ should be adjusted to minimize the locational cost function $\mathcal{H}$. According to \cite{Cortes_04}, a necessary condition for optimality is:

\begin{equation} \label{E6}
\boldsymbol{p}_i(t) = \boldsymbol{c}_i(\boldsymbol{p}, t), \quad i \in \mathcal{N},
\end{equation}
where $\boldsymbol{c}_i$ is the centroid of the Voronoi cell $V_i$ computed with respect to the density function as defined in \cite[Eq.~(3)]{Bai2022}.

\begin{remark}
\label{optimal}
When condition~\eqref{E6} is satisfied, the multi-agent system reaches a locally optimal configuration, where each agent is located at the centroid of its corresponding Voronoi cell. This configuration defines a CVT, which minimizes the locational cost function under the given density distribution.
\end{remark}

\section{Cooperative transportation strategy and controller design}
 \label{Controller design}
\subsection{Problem statement}
In this work, we consider a multi-agent cooperative transportation problem in a bounded workspace. A team of $N$ agents moves with limited sensing range and can communicate only with nearby neighbors.
Each agent is modeled as a first-order integrator:
\begin{equation}\label{dynamics}
    \dot{\boldsymbol{p}}_{i}(t) = \boldsymbol{u}_{i}(t),
\end{equation}
where $\boldsymbol{u}_i \in \mathbb{R}^2$ denotes the control input of agent $i$. The environment contains multiple obstacles (objects) whose number, positions, sizes, and shapes are all unknown a priori. Each agent can only obtain local information about nearby objects within its sensing range, and no centralized assignment is available.

The objective is to design a control strategy that enables the agents to (i) detect and localize objects using only local sensing, (ii) autonomously and adaptively self-group into multiple teams, where the team size increases with the object scale (e.g., larger objects attract more agents), and (iii) cooperatively transport each object to a given destination while maintaining safe coordination.

\subsection{Control strategy design}
To realize the control objective, we present in this section the overall framework for multi-agent object detection and cooperative transportation. First, a CVT-based strategy is introduced to guide agents toward detected objects. Then, a CBF-based controller is developed to ensure efficient spatial organization and safe cooperative interactions.

\subsubsection{Object detection and caging formation}
The problem of detecting and interacting with unknown objects can be formulated as a multi-agent coordination problem, in which agents operate within a workspace and autonomously respond to locally perceived object-related information. A key challenge in this scenario is that the spatial influence of target objects is initially unknown to the agents and must be inferred online through local sensing.

To address this challenge, we introduce a time-varying density function, denoted by $\phi(\boldsymbol{q}, t)$. By employing the proposed density function, agents can adaptively encode object-induced spatial importance based on local observations, which in turn guides the self-organized distribution and enclosure of agents around detected objects.

The form of $\phi(\boldsymbol{q}, t)$ is as follows
\begin{align}
\label{eq6}
\phi(\boldsymbol{q}, t) =\ & \phi_0 + \sum_{i=1}^{N} \rho_i(t) \frac{\omega_i}{2\pi \sigma_{x_i} \sigma_{y_i}} \notag \\
& \exp\left( -\frac{1}{2} \left( 
\frac{(q_x - \mu_{x_i})^2}{\sigma_{x_i}^2} + 
\frac{(q_y - \mu_{y_i})^2}{\sigma_{y_i}^2} 
\right) \right),
\end{align}
where $\phi_{0} > 0$ is a small positive constant that ensures the density function remains strictly positive throughout the domain. For each agent $i \in \mathcal{N}$, the parameters $\omega_i > 0$, $\sigma_{x_i} > 0$, and $\sigma_{y_i} > 0$ represent the weight and the standard deviations of the $i$-th Gaussian component along the $q_x$ and $q_y$ axes, respectively. The mean values $\mu_{x_i}$ and $\mu_{y_i}$ specify the center of the $i$-th Gaussian component, typically corresponding to the current position of agent $i$. The vector $\boldsymbol{q} = [q_x, q_y]^{\top}$ denotes a generic point in the domain $\mathcal{D}$. 

In practice, once a target object is identified through sensor detection, $\boldsymbol{q}$ represents any candidate point in the vicinity of the detected object when evaluating the density function and computing Voronoi centroids. The coordinates $(q_x, q_y)$ are sampled from the object-induced region during numerical integration, and the density peaks are centered at $(\mu_{x_i}, \mu_{y_i})$ corresponding to the positions of agents that have detected the object. This formulation enables the CVT computation to distribute agents according to the actual spatial extent and geometry of the detected object, rather than forcing them to converge to a single geometric center, thereby facilitating balanced enclosure and cooperative transportation of objects with arbitrary shapes.

The binary switching variable $\rho_i(t) \in \{0, 1\}$ indicates whether agent $i$ currently detects a significant region. It is defined as
\begin{equation}
\label{eq:rho}
\rho_i(t) =
\begin{cases}
1, & \text{if agent $i$ detects a target object at time $t$}, \\
0, & \text{otherwise}.
\end{cases}
\end{equation}
This detection-dependent activation mechanism dynamically modulates the density function, enabling the multi-agent system to adapt its spatial organization in response to detected objects. With the proposed event-triggered density function, the overall procedure unfolds as follows, from the initial deployment of agents to the formation of cooperative enclosure structures around target objects.

At the beginning, agents are uniformly distributed across the workspace, without any prior knowledge of the location, size, or shape of target objects. If no agent detects an object within its sensor range, the agents gradually increase their mutual distances, leading to a more dispersed formation that facilitates efficient exploration of the workspace.

When an agent detects a target object within its sensor range, the detection status $\rho_i(t)$ in Equation~\eqref{eq:rho} is set to $1$, which updates the density function $\phi(\boldsymbol{q}, t)$ defined in Equation~(4). Each peak of the density corresponds to the current position of an agent that has detected the object, thereby encoding object-induced spatial importance.

Using the updated density, a Voronoi tessellation is computed as in Equation~(1). Each agent then calculates the centroid of its Voronoi region with respect to the current density, following Equation~(4). By minimizing the locational cost function defined in Equation~(2), agents are guided toward spatial configurations that promote balanced distribution around detected objects.

As agents move toward their respective centroids, additional agents may detect the object. New density peaks are added at their positions, resulting in a multimodal density distribution that reflects the evolving set of object-detecting agents. Through this iterative process, the system gradually converges to a cooperative enclosure configuration, in which agents are spatially organized around the object in a balanced manner, providing a suitable structure for subsequent cooperative transportation.

\subsubsection{CBF-based controller design}

Here we presents the detailed formulation of the proposed composite motion controller, which integrates CVT-based spatial organization with a CBF-based constraint mechanism. To achieve effective and safe cooperative enclosure and transportation of target objects, we design a composite motion controller that combines CVT-based positioning for balanced agent distribution with CBFs to enforce safety and structural constraints, including collision avoidance and the prevention of undesired agent clustering during close-proximity cooperation.

Following the CVT principle, the nominal control input is defined as:
\begin{equation}
    \boldsymbol{u}_{i}^{\mathrm{nom}}(t) = -k(\boldsymbol{p}_{i}(t) - \boldsymbol{c}_{i}(t)),
\end{equation}
where $\boldsymbol{c}_i(t)$ is the centroid of the Voronoi region $V_i$ computed with respect to a density function $\phi(\boldsymbol{q}, t)$~\cite{Cortes_04}, and $k > 0$ is a positive gain.

To guarantee safety and regulate close-proximity interactions among agents, we incorporate CBFs into the control design. For each pair of agents $(i,j)\in \mathcal{N} \times \mathcal{N}$, $i\ne j$, we define the following safety set:
\begin{equation}
\begin{aligned}
\mathcal{C}_{ij}(t) = \bigg\{\, i \in \mathcal{N} \ \bigg|\ 
& \|\boldsymbol{p}_i(t) - \boldsymbol{p}_j(t)\|^2 
  - d_{\mathrm{min}}^2 \geq 0,  
\\
& \qquad\qquad\text{for some } j \ne i \,\bigg\}
\end{aligned}
\end{equation}
where $d_{\mathrm{min}}\in \mathbb{R}$ is the minimum safety distance.

A zeroing CBF $h_{ij}(t) = \|\boldsymbol{p}_i(t) - \boldsymbol{p}_j(t)\|^2 - d_{\mathrm{min}}^2$  (see~\cite{Ames2017}) ensures that the constraint $h_{ij}(t) \geq 0$ is forward invariant if the following condition is satisfied:
\begin{equation}
    \dot{h}_{ij}(t) + \gamma h_{ij}(t) \geq 0,
\end{equation}
 for all $t\geq 0$, where $\gamma > 0$ is a positive constant. This condition is enforced by solving a quadratic program (QP) at each time step:

\begin{align}
    \min_{\boldsymbol{u}_i} \quad & \|\boldsymbol{u}_i(t) - \boldsymbol{u}_i^{\mathrm{nom}}(t)\|^2 \notag \\
    \text{s.t.} \quad & \frac{\partial h_{ij}(t)}{\partial \boldsymbol{p}_i(t)} \boldsymbol{u}_i(t) 
    + \frac{\partial h_{ij}(t)}{\partial \boldsymbol{p}_j(t)} \boldsymbol{u}_j(t) \notag  + \gamma h_{ij}(t)\\
    & \geq 0, \quad
    \text{for all } (i,j) \in \mathcal{N} \times \mathcal{N},\ i \ne j.
\end{align}

Through this formulation, agents dynamically update their positions and local sensing information, and the Voronoi partitions and density functions evolve accordingly, enabling a safe and balanced spatial organization around detected objects.
The proposed composite controller integrates CVT-based spatial organization with CBF-based constraint enforcement, providing both coordination and safety guarantees for cooperative enclosure and transportation tasks.

\section{Simulation results}\label{simulation}

\begin{figure}[htbp]
    \centering
    \subfloat[]{\includegraphics[width=0.50\textwidth]{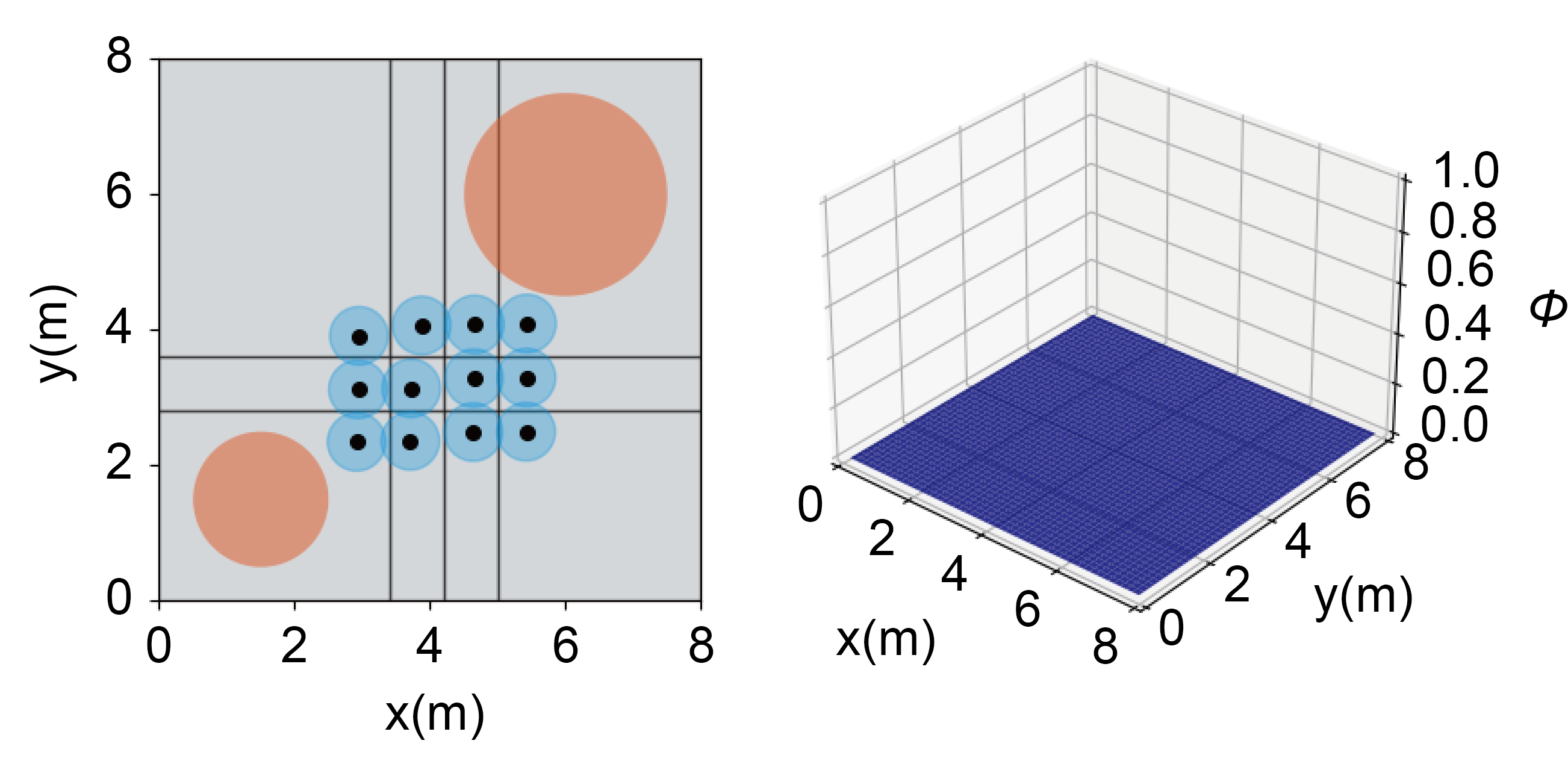}}
    \hfill
    \subfloat[]{\includegraphics[width=0.50\textwidth]{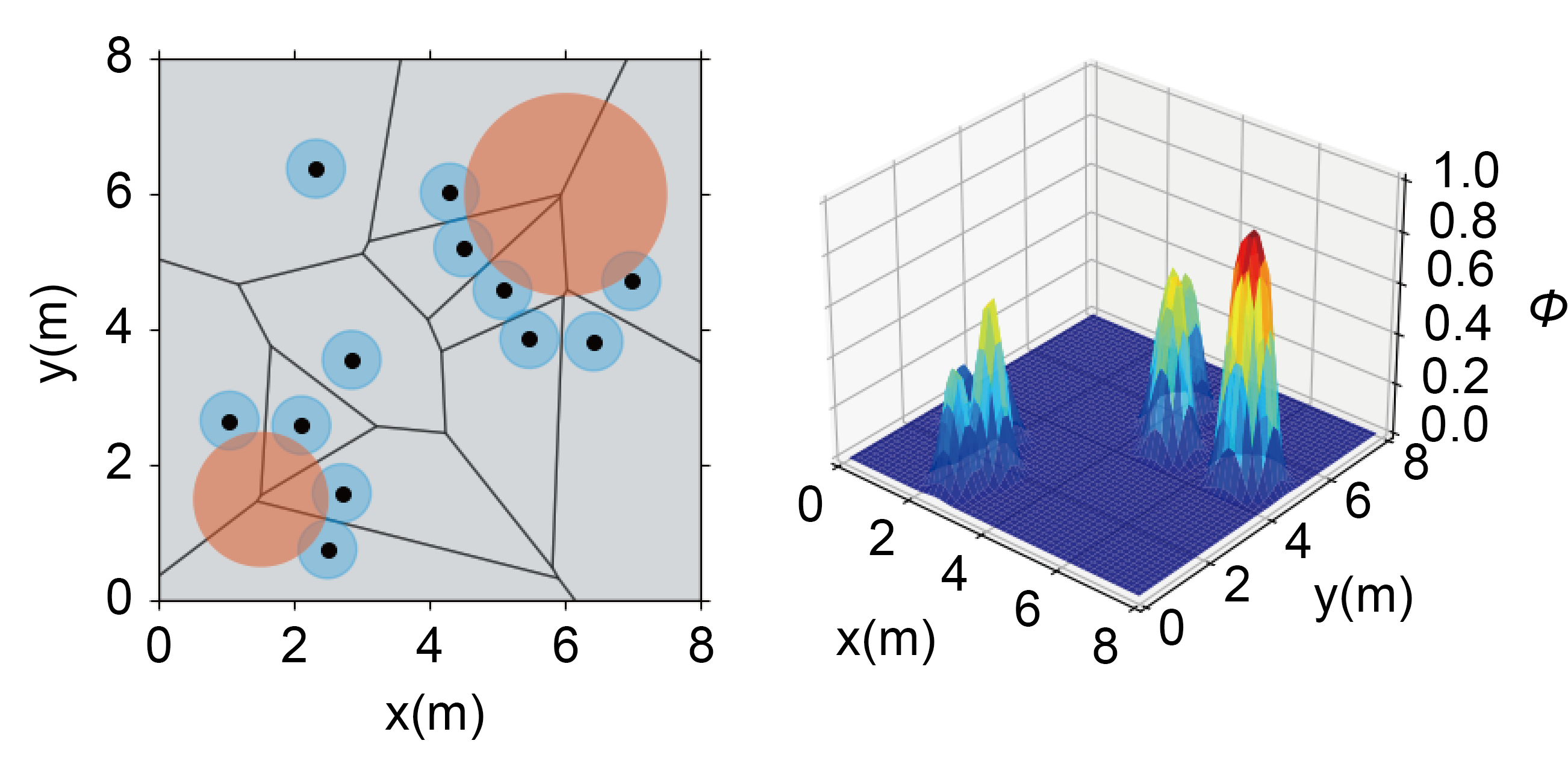}}\\
    \subfloat[]{\includegraphics[width=0.50\textwidth]{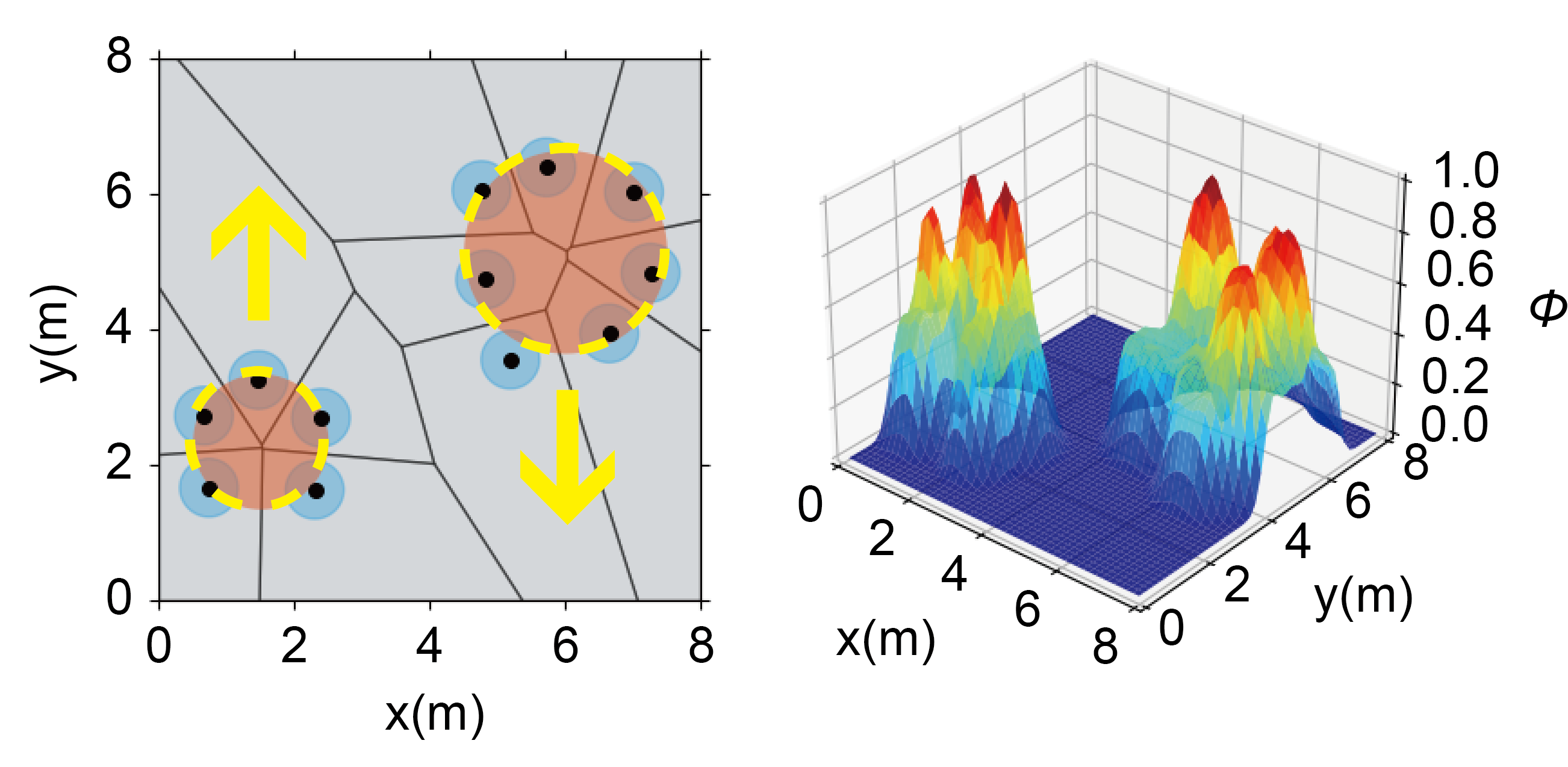}}
    \hfill
    \subfloat[]{\includegraphics[width=0.50\textwidth]{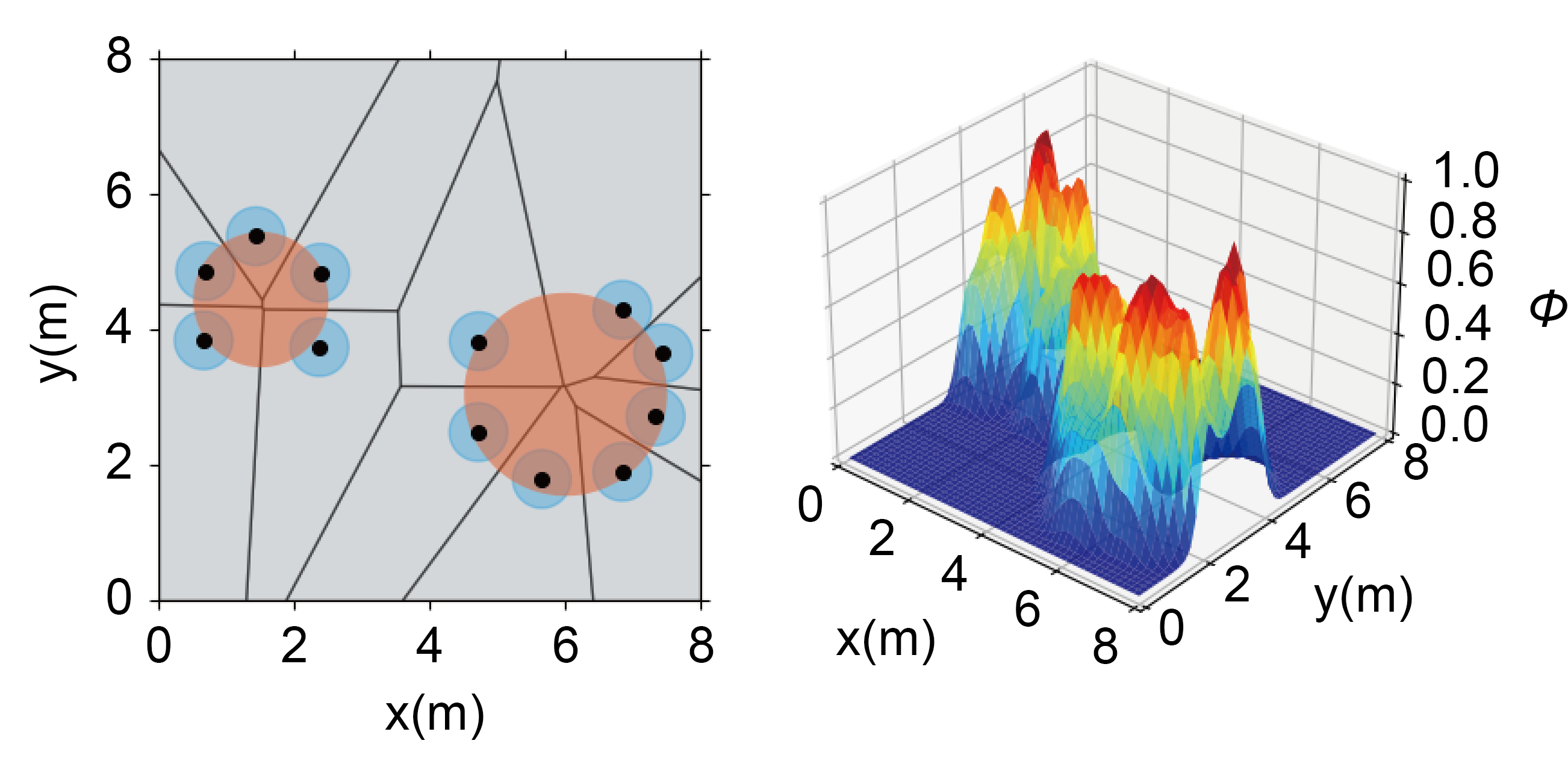}}
    \caption{Snapshots of the simulation process at different time steps: (a) $t = 0\,\mathrm{s}$, (b) $t = 31\,\mathrm{s}$, (c) $t = 82\,\mathrm{s}$, and (d) $t = 124\,\mathrm{s}$. 
In each snapshot, the left panel illustrates the spatial configuration of the agents and objects in the workspace, where black dots denote the agents and blue circles indicate their local detection ranges. The orange circles represent target objects with different sizes. The right panel shows the corresponding object-induced density function $\phi(\boldsymbol{q}, t)$, which is dynamically constructed based on agents’ local detections and guides the cooperative enclosure behavior.
}
    \label{fig:coverage_snapshots_2AOI}
\end{figure}

\begin{figure}[htbp]
    \centering
    \includegraphics[width=0.45\textwidth]{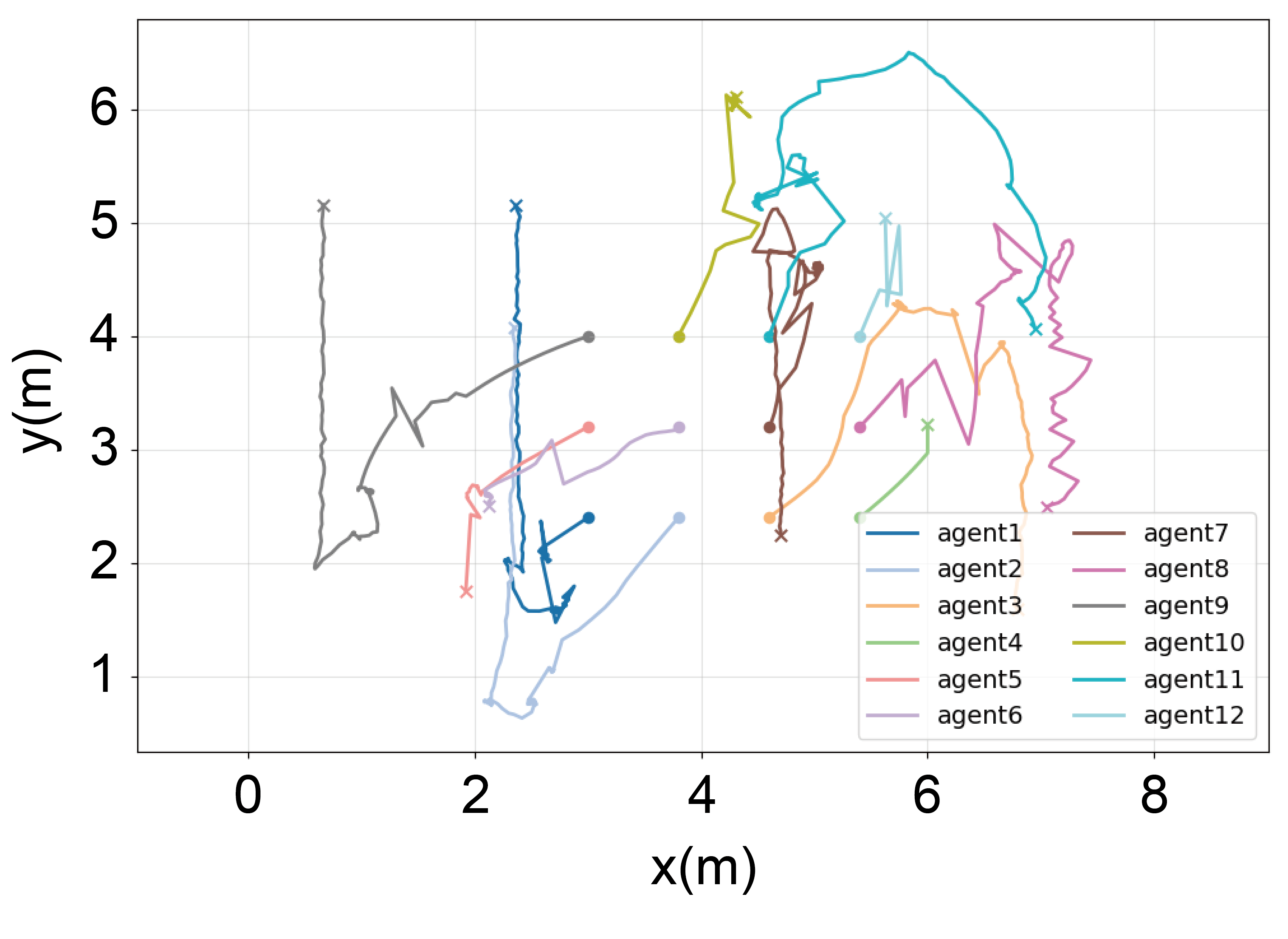}
    \caption{Agent trajectories during the object detection and transportation process.}
    \label{fig:agent_paths_2AOI}
\end{figure}

To evaluate the effectiveness of the proposed multi-agent cooperative transportation framework, numerical simulations were conducted in a bounded two-dimensional workspace. In the simulation environment, the orange circular regions represent cargos to be cooperatively transported by the multi-agent system. The cargos have different sizes, corresponding to different transportation demands. At the beginning of the simulation, all agents are initialized in a compact formation near the center of the workspace, without any prior knowledge of the cargos’ locations or sizes. Each agent is equipped with a limited sensing range, illustrated by blue circles, while black dots indicate the agents’ positions. The Voronoi partitions induced by the agent configuration are shown in the left panels of Fig.~\ref{fig:coverage_snapshots_2AOI}.

Fig.~\ref{fig:coverage_snapshots_2AOI}(a) illustrates the initial configuration of the system. At this stage, none of the agents has detected any cargo, and the density function remains inactive, as indicated by the flat surface in the right panel. The agents gradually disperse from the initial compact formation to explore the workspace more effectively.
Once one or more agents detect a cargo within their sensing range, object-induced density functions are generated at the detecting agents’ positions. Fig.~\ref{fig:coverage_snapshots_2AOI}(b) shows this detection stage, where agents begin to respond to the presence of cargos. 
As the simulation progresses, additional agents are attracted toward the cargos and independently generate density functions. These density functions are superimposed, forming multiple peaks in the density distribution, as shown in Fig.~\ref{fig:coverage_snapshots_2AOI}(c). Through this process, agents self-organize into enclosure formations around each cargo and initiate cooperative transportation. Due to the size-dependent density generation, the smaller cargo attracts fewer agents, whereas the larger cargo attracts a greater number of agents, enabling adaptive and decentralized agent allocation without explicit task assignment. 
In the snapshot, the yellow arrows indicate the directions for the transport of the cargos. Specifically, the left cargo is cooperatively transported upward in the vertical direction by five agents approaching from the lower-left side, while the right cargo is transported downward by seven agents approaching from the upper-right side.
Fig.~\ref{fig:coverage_snapshots_2AOI}(d) depicts a later stage of the simulation, where the cargos are being transported toward the workspace boundaries. As the agents approach the boundary, their transportation velocities are gradually reduced, and the system smoothly transitions to a stopping behavior. Throughout the transportation process, the Voronoi partitions continuously adapt to the evolving agent configurations, supporting balanced spatial organization around each cargo.
Fig.~\ref{fig:agent_paths_2AOI} presents the trajectories of all agents over the entire simulation. The trajectories demonstrate smooth and coordinated motions toward the cargos, followed by stable enclosure and transportation behaviors. No excessive oscillations or undesired aggregation are observed, highlighting the robustness and stability of the proposed framework.

Overall, the simulation results demonstrate that the proposed method enables multiple agents to autonomously detect unknown cargos of different sizes, adaptively form cooperative enclosure structures, and perform stable and decentralized cooperative transportation. The integration of size-dependent density generation, CVT-based spatial organization, and CBF-based structural regulation allows the system to achieve flexible and scalable multi-agent cooperative transportation in environments with multiple heterogeneous cargos.

\section{Conclusion} \label{conclusion}
This paper proposed a density-driven cooperative transportation framework that enabled multi-agent systems to detect and transport multiple cargos without prior knowledge of their size, shape, or location, using only local sensing and neighbor interactions. Object-induced density functions generated attraction fields for autonomous recruitment and adaptive team formation, while CVT-based design provided balanced spatial organization and naturally allocates more agents to larger cargos. A CBF-based design was further incorporated to enforce safety and structural constraints, preventing clustering and promoting stable, symmetric enclosure during close-proximity transport. Simulations validated the effectiveness of the proposed approach under unknown cargo conditions. Future work will incorporate more realistic cargo properties (e.g., weight) and validate the framework in physical multi-robot experiments.

\addtolength{\textheight}{-12cm}   





\end{document}